\documentclass[preprint,pra,aps,graphicx,multicol,epsfig,showpacs]{revtex4}
\usepackage{graphics}

\begin{document}

\title{Stabilization and destabilization of second-order solitons against
perturbations in the nonlinear Schr\"{o}dinger equation.}
\author{Hilla Yanay$^{1}$, Lev Khaykovich$^{1}$ and Boris A. Malomed$^{2}$}
\affiliation{$^{1} $Department of Physics, Bar-Ilan University, Ramat-Gan, 52900 Israel,}
\affiliation{$^{2}$Department of Physical Electronics, School of Electrical Engineering,
Faculty of Engineering, Tel Aviv University, Tel Aviv 69978, Israel}

\begin{abstract}
We consider splitting and stabilization of second-order solitons (2-soliton
breathers) in a model based on the nonlinear Schr\"{o}dinger equation
(NLSE), which includes a small quintic term, and weak resonant nonlinearity
management (NLM), i.e., time-periodic modulation of the cubic coefficient,
at the frequency close to that of shape oscillations of the 2-soliton. The
model applies to the light propagation in media with cubic-quintic optical
nonlinearities and periodic alternation of linear loss and gain, and to BEC,
with the self-focusing quintic term accounting for the weak deviation of the
dynamics from one-dimensionality, while the NLM can be induced by means of
the Feshbach resonance. We propose an explanation to the effect of the
resonant splitting of the 2-soliton under the action of the NLM. Then, using
systematic simulations and an analytical approach, we conclude that the weak
quintic nonlinearity with the self-focusing sign stabilizes the 2-soliton,
while the self-defocusing quintic nonlinearity accelerates its splitting. It
is also shown that the quintic term with the self-defocusing/focusing sign
makes the resonant response of the 2-soliton to the NLM essentially broader,
in terms of the frequency.
\end{abstract}

\pacs{03.75.Lm; 42.65.Tg; 42.81.Dp; 05.45.Yv}
\maketitle

\textbf{A well-known consequence of the integrability of the one-dimensional
nonlinear Schr\"{o}dinger equation with the self-focusing cubic nonlinear
term is that this equation gives rise to an infinite family of }$N$\textbf{%
-th order soliton solutions, in the form of periodically oscillating
breathers (the oscillation period is the same for all }$N\geq 2$\textbf{).
These solutions may be interpreted as bound states of }$N$\textbf{\
overlapping fundamental solitons with different amplitudes. Their binding
energy being exactly zero, all }$N$\textbf{-soliton complexes are unstable
against perturbations of the initial conditions, which can split them into
slowly separating constituent fundamental pulses. In this work, we aim to
study the stabilization/destabilization of the $2$- (second-order) solitons
by a weak additional quintic self-focusing/self-defocusing term. Various
origins of such additional nonlinearity, of either sign, are known in
Bose-Einstein condensates (BECs) and nonlinear optics, being within the
reach of available experimental techniques. We argue that our predictions of
the change in the stability of the higher-order solitons, and of a
possibility to implement the precise control of the soliton dynamics by
means of the quintic nonlinear term, may be realized in experiments with
optical and matter-wave solitons. The analysis is performed using a
combination of direct simulations and analytical approximations.}

\section{Introduction}

The control of soliton dynamics has been drawing a great deal of interest
both as a fundamental problem and a topic with a vast spectrum of
applications -- in particular, to optics and matter waves \cite{management}.
It is well known that the nonlinear Schr\"{o}dinger equation (NLSE), being
an integrable one, supports an infinite sequence of exact higher-order
soliton solutions, which may be understood as bound states of strongly
overlapping fundamental solitons \cite{TheorySolitons}. However, within the
framework of the integrable equation, the binding energy of the exact
multi-soliton complex is always equal to zero \cite{TheorySolitons}, hence
the higher-order solitons are unprotected (unstable) against perturbations
of initial conditions, which can induce splitting into fundamental
constituents. For example, the second- and third-order solitons (which are
often briefly called 2-soliton and 3-soliton, respectively) readily split
into sets of two and three fundamental solitons, with amplitude ratios $1:3$
and $1:3:5$.

Different approaches were proposed to stimulate the splitting and make it a
real physical effect. One possibility is to introduce a specific nonlinear
dissipation into the model by adding nonconservative terms to the NLSE, such
as the one accounting for the intrapulse Raman scattering in optical fibers
\cite{SolitonsInOC}. Other physically relevant settings are those with a
periodic modulation of either the group-velocity-dispersion (GVD) or
nonlinearity coefficient in the NLSE, which is known as the dispersion
management (DM) and nonlinearity management (NLM), respectively \cite%
{management}. The DM in optical fibers can be of both sign-alternating and
sign-preserving types (in the former case, the GVD coefficient periodically
changes between positive and negative, alias normal-GVD and anomalous-GVD,
values). The format of the DM can be piecewise-constant, built as an
alternation of fiber segments with different values of the GVD coefficient,
or sinusoidal. It was predicted \ that the latter format, with a mild
modulation amplitude that does not imply the change of the sign of the GVD,
can induce the splitting of both fundamental \cite{Scripta} and higher-order
solitons \cite{Malomed93,Bauer95}. Both effects have been experimentally
demonstrated in Ref. \cite{Sysoliatin08}, which made use of a specially
fabricated fiber with the diameter subjected to the sinusoidal modulation
along the fiber, thus inducing the modulation of the local GVD coefficient.

Various effects of the NLM in the context of fiber-optic telecommunications
were theoretically studied in Refs. \cite{Hasegawa91,Pare,Radik}, and the
integration of the NLM with the DM was considered in Ref. \cite{Radik2}. In
terms of Bose-Einstein condensates (BECs) in dilute atomic gases, which, in
the mean-field approximation, is also described by the NLSE (called the
Gross-Pitaevskii equation, in that context \cite{GPE}), the NLM represents
the application of the Feshbach-resonance technique to the BEC in the case
when the resonance is induced \ by a variable (ac) magnetic field \cite%
{Greece}. In the latter case, Ref. \cite{Sakaguchi04} has demonstrated that
a small-amplitude variable part of the nonlinearity coefficient gives rise
to resonant splitting of higher-order solitons into fundamental ones,
provided that the NLM frequency is close to the frequency of free shape
oscillations of the higher-order bound state.

In this work we demonstrate that the addition of a very weak
\textit{quintic nonlinearity} dramatically changes the stability of
2-solitons under the action of the NLM. Namely, the 2-soliton's
splitting time becomes significantly larger (smaller) in the
presence of weak additional attraction (repulsion), represented by a
self-focusing (defocusing) quintic term. These predictions may be
tested experimentally in nonlinear optics and BEC. A challenging
possibility is to create effectively stable 2-solitons using a weak
self-focusing (attractive) quintic nonlinearity, and control their
behavior by means of the weak NLM. While both the small quintic term
and weak NLM represent perturbations that break the integrability of
the underlying NLSE, the stabilization of the 2-soliton states under
the combined action of \emph{both perturbations} implies an
intriguing possibility of their mutual compensation, leading to an
effective extension of the quasi-integrable behavior of the
solitons.

In optics, the cubic-quintic (CQ) nonlinearity with different sign
combinations of the two terms were predicted \cite{18} and observed \cite{19}
in aqueous colloids, as well as in dye solutions \cite{20} and, very
recently, in thin ferroelectric films \cite{ferro}. The same nonlinearity
was also predicted, via the cascading mechanism, in two-level media \cite{21}%
. The self-defocusing quintic nonlinearity, accounted for by a proximity to
the resonant two-photon absorption, was also observed in other optical
materials \cite{22}.

In the description of effectively one-dimensional BEC settings, a universal
\emph{self-attractive} quintic term in the respective Gross-Pitaevskii
equation (as said above, the attractive quintic nonlinearity should
facilitate the creation of stabilized 2-solitons)\ accounts for the
deviation from the exact one-dimensionality, i.e., a finite transverse size
of the corresponding trap for the atomic condensate \cite%
{Shlyap02,Luca,Khaykovich06}. Besides that, a self-defocusing quintic term
may take into regard three-body collisions in the BEC, provided that the
related losses are negligible \cite{Fatkhulla}.

As concerns the time-periodic modulation of the cubic coefficient,
dealt with in this work, in optics it may naturally arise as a
result of the periodic alternation of the linear loss and
compensating gain (a well-known transformation removes the
respective linear terms in the NLSE, mapping them into the effective
NLM) \cite{Hasegawa91}. As mentioned above, the same modulation in
BEC represents the action of the Feshbach resonance controlled by
the ac magnetic field. Thus, both the CQ nonlinearity and NLM are
generic features of numerous physical settings.

The paper is organized as follows. Results obtained by means of systematic
simulations, that demonstrate the stabilization/destabilization of the
2-soliton, subject to the action of the resonant or near-resonant NLM, under
the action of the weak quintic term, that corresponds, respectively, to the
self-attraction/repulsion, are reported in Section II. Analytical
approximations, which make it possible to explain the underlying effect of
the resonant splitting of higher-order solitons under the action of the weak
NLM, and also the stabilization/destabilization of the 2-soliton, induced by
the quintic term, are presented in Section III. These approximations are
based on analysis of the system's energy in the presence of the NLM and
quintic term. Finally, Section IV concludes the paper.

\section{Splitting of the second-order solitons}

\subsection{Second-order soliton in nonlinear Schr\"{o}dinger equation with
cubic-quintic nonlinearity}

In this work, we take the one-dimensional NLSE for wave function $\phi
\left( x,t\right) $ in the usual scaled form,
\begin{equation}
i\frac{\partial \phi }{\partial t}=-\frac{1}{2}\frac{\partial ^{2}\phi }{%
\partial x^{2}}-g|\phi |^{2}\phi +\epsilon |\phi |^{4}\phi ,
\label{Eq_NLSEquintic}
\end{equation}%
where $g>0$ is the coefficient accounting for the cubic self-attraction,
which is set to be $g\equiv 1$ in the absence of the NLM. In physically
relevant realizations of Eq. (\ref{Eq_NLSEquintic}), the dimensionless
quintic-interaction constant, which accounts for the higher-order
self-attraction/repulsion in case of $\epsilon <0$/$\epsilon >0$, is a small
parameter, $|\epsilon |\ll 1$. Energy $E$ (the Hamiltonian), from which Eq. (%
\ref{Eq_NLSEquintic}) can be derived as $i\partial \phi /\partial t=\delta
E/\delta \phi ^{\ast }$, where $\delta /\delta \phi ^{\ast }$ stands for the
variational derivative with respect to the complex-conjugate field \cite%
{TheorySolitons}, is%
\begin{equation}
E=\int_{-\infty }^{+\infty }\left( \frac{1}{2}\left\vert \frac{\partial \phi
}{\partial x}\right\vert ^{2}-\frac{g}{2}|\phi |^{4}+\frac{\epsilon }{3}%
|\phi |^{6}\right) dx.  \label{E}
\end{equation}

\begin{figure}[tbp]
{\centering\resizebox*{0.45\textwidth}{0.45\textheight} {{%
\includegraphics{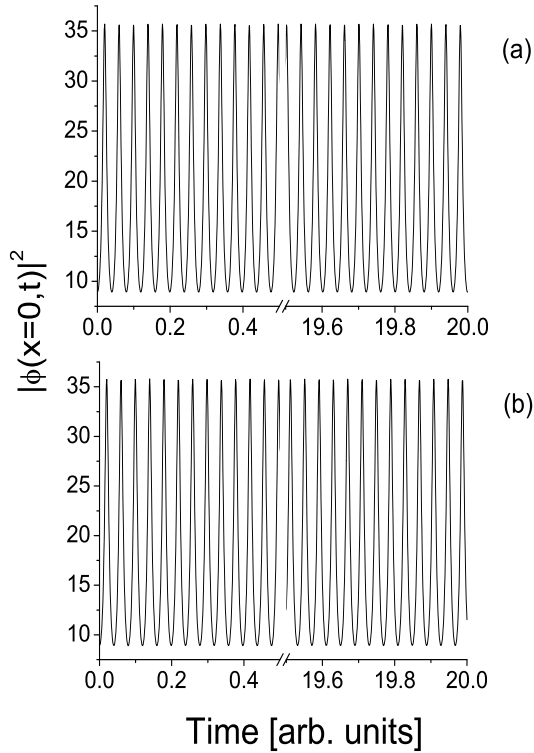}}} }
\caption{Oscillations of the peak density of the second-order soliton in the
absence of the quintic nonlinearity, $\protect\epsilon =0$, (a), and in the
presence of a weak attractive quintic nonlinearity, with $\protect\epsilon %
=-1.6\times 10^{-3}$, (b).}
\label{SecondSolitonOscillations}
\end{figure}

Exact fundamental-soliton solutions to Eq. (\ref{Eq_NLSEquintic}) are known
for either sign of $\epsilon $ \cite{Pushkarov79,Khaykovich06}. In the case
of the integrable cubic NLSE, with $\epsilon =0$, exact $n$-soliton
solutions are generated by initial conditions
\begin{equation}
\phi (x,t=0)=nA~\mathrm{sech}(Ax),  \label{Eq_InitialCondition}
\end{equation}%
with integer $n\geq 1$ \cite{Satsuma74}. For all $n\geq 2$, they are
breathers whose shape oscillates with the frequency independent of $n$,
\begin{equation}
\omega _{\mathrm{sh}}=4A^{2}  \label{Eq_OscillationFrequency}
\end{equation}%
($2\pi /\omega _{\mathrm{sh}}$ is usually called the soliton period). The
explicit solution for the 2-soliton is relatively simple:%
\begin{equation}
\phi (x,t)=4Ae^{iA^{2}t/2}\frac{\cosh \left( 3Ax\right) +3e^{4iA^{2}t}\cosh
\left( Ax\right) }{\cosh \left( 4Ax\right) +4\cosh \left( 2Ax\right) +3\cos
\left( 4A^{2}t\right) }.  \label{exact}
\end{equation}%
The energy of the 2-solution, taken as per Eq. (\ref{E}) with $g=1$ and $%
\epsilon =0$, is%
\begin{equation}
\left( E_{0}\right) _{\mathrm{2-sol}}=-\left( 28/3\right) A^{3}  \label{E0}
\end{equation}%
(as said above, it is identical to the sum of the energies of two
far separated fundamental solitons with amplitudes $3A$ and $A$ into
which the 2-soliton complex may split). Figure
\ref{SecondSolitonOscillations}(a) displays oscillations of the peak
density $|\phi (x=0,t)|^{2}$ of the 2-soliton, generated by initial
condition (\ref{Eq_InitialCondition}) with $n=2 $.

Numerical simulations of Eq. (\ref{Eq_NLSEquintic}) are performed using the
split-step Fourier spectral method with 4096 Fourier modes and absorptive
boundary conditions. Figure \ref{SecondSolitonOscillations}(b) displays $504$
oscillations of the peak density of the numerical solution with a weak
attractive quintic nonlinearity, corresponding to $\epsilon =-1.6\times
10^{-3}$. Despite the absence of exact solutions for higher-order solitons
at $\epsilon \neq 0$, the oscillations persist indefinitely long without any
tangible decay or distortion, the only difference from the exact solution
presented in panel (a) for $\epsilon =0$ being a small increase in the
oscillation frequency, as a result of the weak additional self-attraction.
In view of the apparent robustness of this solution at small values of $%
\left\vert \epsilon \right\vert $, we will continue to refer to it as the
second-order soliton.

\subsection{Resonant splitting of the second-order soliton}

We introduce the NLM, in the form of the time-periodic (ac) modulation of
the nonlinearity, by setting
\begin{equation}
g(t)=1+b\sin (\omega t),  \label{Eq_Modulation}
\end{equation}%
in Eq. (\ref{Eq_NLSEquintic}), where amplitude $b$ of the perturbation is
small, $|b|\ll 1$, and the modulation frequency is kept in resonance with
the shape oscillations, $\omega =\omega _{\mathrm{sh}}$. Figure \ref%
{Splitting} shows a typical example of the evolution of the wave function
subject to the action of the resonant perturbation for $\epsilon =0$ (with
the cubic nonlinearity only) and $b=5\times 10^{-3}$. The 2-soliton splits
into fundamental solitons with amplitudes related as $1:3$, as expected from
the exact solution available at $b=0$. The respective velocity ratio of the
splinters is $3:1$, in agreement with the prediction based on the momentum
conservation (it follows from the fact that the effective masses of the two
splinters are in the same ratio as their amplitudes \cite{TheorySolitons},
i.e., $1:3$, and the total momentum must remain equal to zero \cite%
{Sakaguchi04}).

\begin{figure}[tbp]
{\centering\resizebox*{0.5\textwidth}{0.22\textheight} {{%
\includegraphics{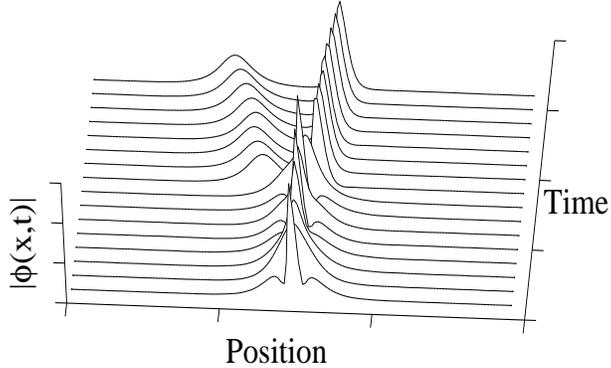}}} }
\caption{A typical example of the splitting of the second-order
soliton under the action of a weak resonant modulation of the
coefficient in front of the cubic nonlinearity, in the absence of
the quintic term.} \label{Splitting}
\end{figure}

The simulations demonstrate that even a very weak resonant perturbation
splits the second-order soliton, but the onset of the splitting requires
time which increases with the decrease of the strength of the ac modulation,
$b$. The smallest perturbation amplitude that we applied in the simulations
was $b=5\times 10^{-5}$. This limitation was determined by the accumulation
of numerical errors and available computational time; however, the results
strongly suggest that there is no cutoff for the resonant splitting, as
concerns the smallness of the perturbation amplitude.

\begin{figure}[tbp]
{\centering\resizebox*{0.5\textwidth}{0.27\textheight} {{%
\includegraphics{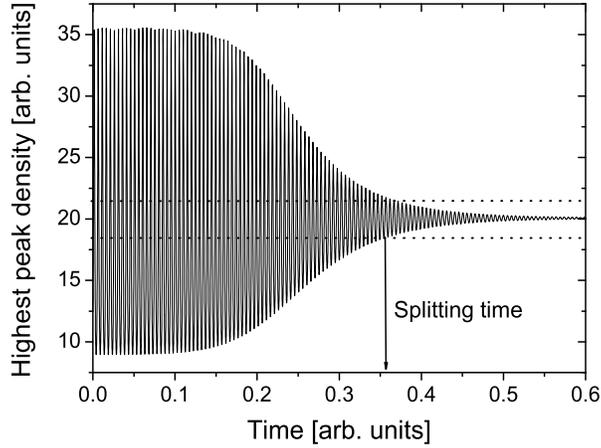}}} }
\caption{A detailed picture of the decay of density oscillations in the
course of the splitting of a second-order soliton into its two fundamental
constituents. Oscillations of the highest peak in the two-soliton waveform
are shown during and after the splitting.}
\label{Splitting-time-definition}
\end{figure}

For the quantitative analysis of the splitting, it is necessary to exactly
define the splitting time, $T$. Figure \ref{Splitting-time-definition} shows
a generic example of the gradual decay of density oscillations of a
second-order soliton in the course of its splitting. Following this picture,
we define $T$ as the time elapsed from the start of the action of the NLM
until the amplitude of the density oscillations drops to $10\%$ of its
initial value in the 2-soliton. By that time, the second-order soliton turns
into well-separated constituents. In fact, adopting another technical
definition of the splitting time produces a little effect on the final
results.

\begin{figure}[tbp]
{\centering\resizebox*{0.48\textwidth}{0.27\textheight} {{%
\includegraphics{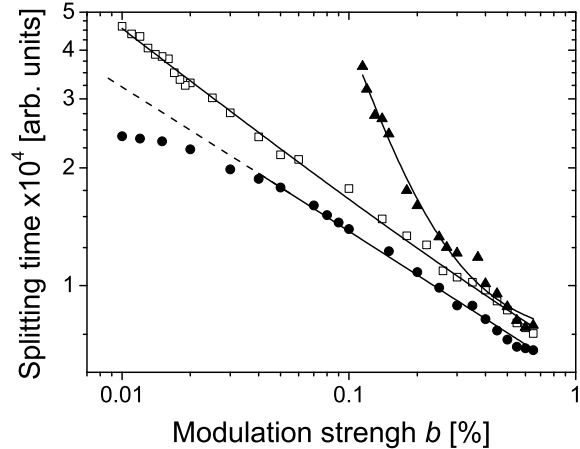}}} }
\caption{The splitting time of the 2-soliton, as produced by the
simulations,versus the ac-drive's amplitude, $b$. Open squares pertain to
the case of the cubic nonlinearity only ($\protect\epsilon =0$), black
triangles are for $\protect\epsilon =-9.1\times 10^{-4}$, and black circles
are for $\protect\epsilon =3.03\times 10^{-4}$. Solid lines show the fit
provided by Eq. (\protect\ref{Eq_FittingFunction}). For positive $\protect%
\epsilon $, the fitting formula (shown by the dashed line in this region)
does not describe the numerical data at smallest values of $b$, which
indicates saturation for extremely weak perturbation strengths.}
\label{SplittingTime}
\end{figure}

The chain of open squares in Fig. \ref{SplittingTime} shows the splitting
time as a function of the perturbation strength, $b$ (in the case of $%
\epsilon =0$), revealing the divergence in the limit of $b\rightarrow 0$. We
fit this dependence to a simple power-law expression,
\begin{equation}
T=a\cdot b^{p}+c  \label{Eq_FittingFunction}
\end{equation}%
where $a,~c$ and $p<0$ are constant parameters and $b$ is taken in
percents. As can be seen in the figure, the fitting formula captures
the behavior of the numerical data well. For $\epsilon =0$, the
fitting parameters are $p=-0.462\pm {0.022}$ and $a=(5.24\pm
{0.59})\times 10^{3}$, $c=(1.44\pm {0.78})\times 10^{3}$.

Further, black triangles and circles in Fig. \ref{SplittingTime} show the
splitting time as a function of the NLM strength in the presence of the weak
quintic nonlinearity of either sign, \textit{viz}., for
\begin{equation}
\epsilon =-9.1\times 10^{-4}~~\mathrm{and}\mathsf{~~}\epsilon =3.03\times
10^{-4},  \label{quintic}
\end{equation}%
respectively. Even this very weak extra nonlinearity affects the
splitting time dramatically: the additional quintic attraction
(repulsion) delays (accelerates) the splitting, thus effectively
stabilizing (destabilizing) the 2-soliton. In particular, the
quintic self-focusing term with the amplitude as small as $\epsilon
=-9.1\times 10^{-4}$ (black triangles in Fig. \ref{SplittingTime})
is enough to increase the spitting time by a factor of $\sim 2.5$
when the ac-drive's strength is $b=1\times 10^{-3}$, as compared to
the case of $\epsilon =0$.

The fit parameters for the data pertaining to the extra quintic attraction
or repulsion, with the values of $\epsilon $ as in Eq. (\ref{quintic}), are,
respectively,
\begin{equation}
p=-1.93\pm {0.16},~~a=(4.2\pm {1.5})\times 10^{2},~~c=(7.25\pm
{0.53})\times 10^{3},  \label{c>0}
\end{equation}%
\begin{equation}
p=-0.372\pm {0.007,~}a=(5.84\pm {0.09})\times 10^{3},~c=0.
\label{c=0}
\end{equation}%
It is worth noting the differences in parameter $c$ for both cases. For $%
\epsilon <0$, positive $c$ in the fitting set (\ref{c>0}) means that, even
for a strong perturbation, a \emph{final waiting time} is required to
observe the splitting of the 2-soliton, which is another manifestation of
its stabilization by the quintic self-focusing term. On the contrary, for $%
\epsilon >0$, the best fit actually required to choose $c<0$--in the
parameter region where formula (\ref{Eq_FittingFunction}) with $c<0$
produces $T>0$. Setting $c=0$ in the fitting set (\ref{c=0}) means that the
simulations demonstrate that the 2-soliton starts splitting instantaneously
under the action of the strong resonant perturbation.

Actually, for large perturbation amplitudes ($b>0.02$), the splitting
produces fundamental solitons with the amplitude ratio different from $1:3$,
which may be explained by effects induced by the relatively strong
perturbation on the constituents of the 2-soliton in the course of the
splitting. In the case of the self-defocusing quintic nonlinearity, $%
\epsilon >0$, the splitting time shows saturation for extremely weak
perturbations (see black circles in Fig. \ref{SplittingTime}), i.e., the
splitting time ceases to grow with further weakening of the perturbation. A
plausible explanation to this feature, which demonstrates the fragility of
the second-order soliton in this situation, is the fact that splitting is
spontaneously initiated by the numerical noise. We also note that the
quintic nonlinearity slightly changes the frequency of the shape
oscillations of the 2-soliton, see Fig. \ref{SecondSolitonOscillations}. The
modulation frequency was modified, accordingly, in the simulations, to
maintain the resonance condition for all cases included in Fig. \ref%
{SplittingTime}.

\begin{figure}[tbp]
{\centering\resizebox*{0.48\textwidth}{0.3\textheight} {{%
\includegraphics{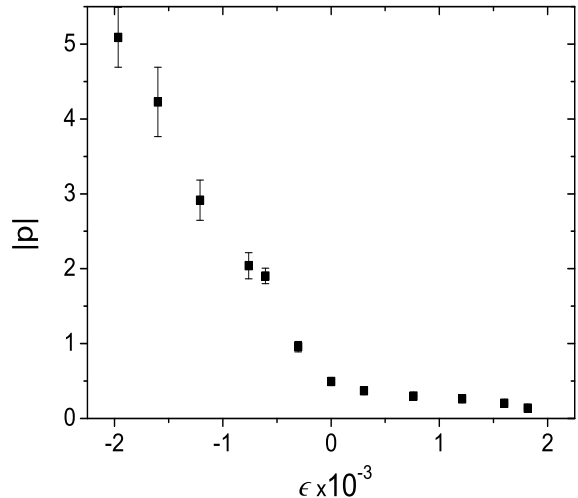}}} }
\caption{The power-law fit parameter, $|p|$, from Eq. (\protect\ref%
{Eq_FittingFunction}), as a function of the strength of the quintic
nonlinearity, $\protect\epsilon $. The error bars represent the error of the
fit to formula (\protect\ref{Eq_FittingFunction}).}
\label{PowerLaw}
\end{figure}

Figure \ref{PowerLaw} summarizes absolute values of exponent $|p|$, i.e.,
the best-fit parameter in Eq. (\ref{Eq_FittingFunction}), as a function of
strength $\epsilon $ of the quintic nonlinearity. The error bars in the
figure represent errors of the fit to the power-law approximation defined as
per Eq. (\ref{Eq_FittingFunction}). For $\epsilon <0$, $|p|$ rapidly
increases with $-\epsilon $, indicating that the second-order soliton
develops strong resistance against the resonant splitting, with the
enhancement of the quintic self-focusing. On the other hand, $|p|$ drops for
$\epsilon >0$, showing the destabilization of the second-order soliton under
the action of the quintic self-defocusing.

\subsection{The near-resonance response}

In Ref. \cite{Sakaguchi04} it was shown that the splitting of the 2-soliton
(in the case of $\epsilon =0$) could also be caused by the temporal
modulation of the coefficient in front of the cubic term with the frequency
slightly different from resonant value (\ref{Eq_OscillationFrequency}).
Here, we aim to confirm this behavior for case of the cubic NLSE and extend
the analysis to the CQ model, with $\epsilon \neq 0$ (in the latter case,
the resonant frequency should be first slightly adjusted, as mentioned
above). For this purpose, we have performed the simulations by varying
modulation strength $b$ at fixed values of the driving frequency and fitting
the so observed splitting time ($T$) to the power-law function, defined as
per Eq. (\ref{Eq_FittingFunction}).

\begin{figure}[tbp]
{\centering\resizebox*{0.48\textwidth}{0.3\textheight} {{%
\includegraphics{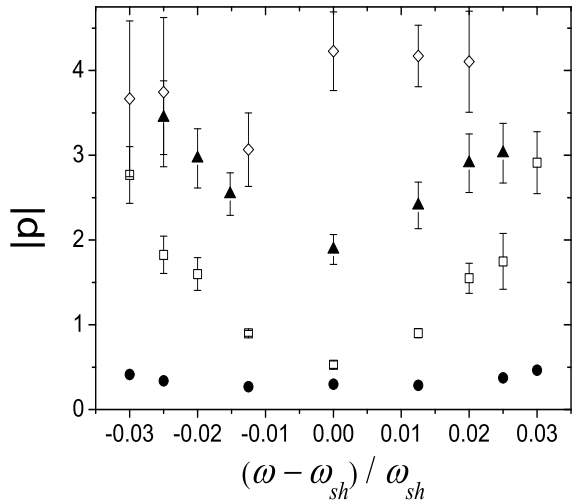}}} }
\caption{The power-law fit parameter, $|p|$, as a function of the modulation
frequency. Open squares, black triangles, open rhombes, and black circles
are, respectively, for the pure cubic nonlinearity ($\protect\epsilon =0$), $%
\protect\epsilon =-7.6\times 10^{-4}$, $\protect\epsilon =-1.6\times 10^{-3}$
(the self-focusing quintic term), and $\protect\epsilon =7.6\times 10^{-4}$
(the self-defocusing one). The error bars represent errors of the fit to the
power-law expression defined as per Eq. (\protect\ref{Eq_FittingFunction}).}
\label{OffResonance}
\end{figure}

Figure \ref{OffResonance} displays the absolute value of the fitting power, $%
|p|$, versus the modulation frequency, for four different values of the
quintic-nonlinearity coefficient, $\epsilon =-1.6\times 10^{-3}$ (open
rhombuses), $\epsilon =-7.6\times 10^{-4}$ (black triangles), $\epsilon =0$
(open squares), and $\epsilon =7.6\times 10^{-4}$ (black circles). As
before, the error bars in the figure represent errors of the fit. Two
noteworthy features are clearly seen in the figure. First, small and large
values of $|p|$ are obtained for the repulsive and attractive quintic
nonlinearity, respectively, in accordance with what was reported above for
the case of the exact resonance. Second, in most cases the presence of the
quintic term of either sign gives rise to \emph{broadening} of the resonant
response, in comparison with the case of the pure cubic nonlinearity
(represented by open squares in Fig. \ref{OffResonance}; the broadening is
weak in the case of the weakest quintic self-focusing, which corresponds to $%
\epsilon =-7.6\times 10^{-4}$, which is represented by the chain of black
triangles). In the case of the repulsive quintic nonlinearity, $\epsilon >0$%
, the dependence shown by the circles in Fig. \ref{OffResonance} is nearly
flat, i.e., the second-order soliton readily splits even at a relatively
large detuning from the resonance. On the contrary, the stabilization of the
2-soliton by the self-focusing quintic nonlinearity is seen to be robust
also under the off-resonance conditions.

\section{Analytical estimates}

A qualitative explanation to some numerical findings reported above can be
provided by an analytical consideration of the model based on Eq. (\ref%
{Eq_NLSEquintic}). First, it is possible to explain the underlying effect of
the resonant splitting of the 2-soliton. Indeed, as mentioned above, the
binding energy of the higher-order solitons is exactly zero in the
integrable NLSE, therefore the splitting may be explained by the fact that
the resonant temporal modulation pumps energy into the bound 2-soliton
state, causing its splitting into constituents, which carry away the excess
energy, in the form of their kinetic energies. In the presence of the small
modulation term in the cubic coefficient given by Eq. (\ref{Eq_Modulation}),
whose frequency is set to coincide with the resonant value (\ref%
{Eq_OscillationFrequency}), the exact evolution equation for energy $E_{0}$
of the unperturbed NLSE, i.e., Eq. (\ref{E}) with $g=1$ and $\epsilon =0$,
can be derived in the following form:%
\begin{equation}
\frac{dE_{0}}{dt}=b\sin \left( 4A^{2}t\right) \int_{-\infty }^{+\infty }%
\mathrm{Im}\left\{ \left( \frac{\partial \phi }{\partial x}\right)
^{2}\left( \phi ^{\ast }\right) ^{2}\right\} dx.  \label{d/dt}
\end{equation}%
In the lowest approximation, one can substitute the unperturbed 2-soliton
solution, as given by Eq. (\ref{exact}), into the right-hand side of Eq. (%
\ref{d/dt}). Under the condition that the shape oscillations of the
2-soliton are synchronized with the temporal modulation in Eq. (\ref%
{Eq_Modulation}) (analysis of of numerical results confirms this conjecture,
in the case of the slowly developing splitting), the time averaging of Eq. (%
\ref{d/dt}) yields an effective \textit{energy-pump rate},
\begin{equation}
\left\langle \frac{dE_{0}}{dt}\right\rangle =\frac{128}{\pi }%
C_{b}A^{5}b\approx ~23.2~A^{5}b,  \label{rate}
\end{equation}%
where the constant is given by the following integral expression$\allowbreak
$,

\begin{eqnarray}
C_{b} &=&\left\vert \int_{0}^{2\pi }d\tau \sin \tau \int_{-\infty }^{+\infty
}d\xi ~\mathrm{Im}\left\{ \left[ \frac{\cosh \left( 3\xi \right) +3e^{-i\tau
}\cosh \xi }{\cosh \left( 4\xi \right) +4\cosh \left( 2\xi \right) +3\cos
\tau }\right] ^{2}\right. \right.   \nonumber \\
&&\times \left. \left. \left[ \frac{\partial }{\partial \xi }\frac{\cosh
\left( 3\xi \right) +3e^{i\tau }\cosh \xi }{\cosh \left( 4\xi \right)
+4\cosh \left( 2\xi \right) +3\cos \tau }\right] ^{2}\right\} \right\vert
~\approx 0.57.  \label{C}
\end{eqnarray}%
This explanation of the gradual onset of the splitting of the 2-soliton
through the pumping of the energy into it by the resonant NLM was not
considered in Ref. \cite{Sakaguchi04}, which was dealing with the
(near-)resonant splitting in the cubic NLSE (with $\epsilon =0$).

The stabilization/destabilization of the 2-soliton by the
self-focusing/defocusing quintic nonlinearity may also be explained by means
of the consideration of the energy. In expression (\ref{E}), the term
corresponding to the quintic nonlinearity in Eq. (\ref{Eq_NLSEquintic})
yields the following expression for the additional energy, obtained by the
substitution of the unperturbed 2-soliton solution (\ref{exact}) and
averaging over the period of its shape oscillations:%
\begin{equation}
\left\langle \Delta E_{\mathrm{2-soliton}}\right\rangle =\left(
2^{11}C_{\epsilon }/3\pi \right) A^{5}\epsilon \approx ~39.5~A^{5}\epsilon ,
\label{DeltaE}
\end{equation}%
with constant%
\[
C_{\epsilon }=\int_{0}^{2\pi }d\tau \int_{-\infty }^{+\infty }d\xi ~\frac{%
\left\{ \left[ \cosh \left( 3\xi \right) +3\left( \cos \tau \right) \cosh
\xi \right] ^{2}+\left( \sin ^{2}\tau \right) \cosh ^{2}\xi \right\} ^{3}}{%
\left[ \cosh \left( 4\xi \right) +4\cosh \left( 2\xi \right) +3\cos \tau %
\right] ^{6}}~\approx 0.18.
\]

Finally, the binding energy of the 2-soliton induced by the quintic term can
be defined as the difference between the respective extra energies of the
pair of free fundamental solitons, with amplitudes $3A$ and $A$, into which
the 2-soliton splits, $\Delta E_{3A+A}=\left( 3^{5}+1\right) (16/45)\epsilon
A^{5}\approx 86.8~\epsilon A^{5}$, and energy (\ref{DeltaE}) of the unsplit
bound state:%
\begin{equation}
\Delta E=\Delta E_{3A+A}-\Delta E_{\mathrm{2-soliton}}\approx 47.3~\epsilon
A^{5}.  \label{binding}
\end{equation}%
The negativeness and positiveness of expression (\ref{binding}) for $%
\epsilon <0$ and $\epsilon >0$, respectively, explains the
stabilization/destabilization of the 2-soliton by the
self-focusing/defocusing quintic nonlinearity.

According to Eq. (\ref{rate}), under the action of the resonant NLM the
energy of the resonantly driven 2-soliton grows, on the average, linearly in
time:
\begin{equation}
E_{0}(t)\approx E_{0}(0)+~23.2~A^{5}bt,  \label{E(t)}
\end{equation}%
where the initial value, $E_{0}(0)$, is energy (\ref{E0}) of the
unperturbed 2-soliton. If all the energy pumped by the NLM drive
would be accumulated in the form of the ``splitting potential",
then, in the case of $\epsilon <0$, the onset of the splitting might
be expected at the moment of time when this stored ``potential"
becomes equal to the binding energy, which would give the splitting
time as $T=\Delta E/\left\langle dE_{0}/dt\right\rangle $. In its
literal form, this result implies $p=-1$ in Eq.
(\ref{Eq_FittingFunction}). The deviation of actual values of $|p|$
from $1,$ as seen in Fig. \ref{PowerLaw}, is explained by the fact
that, within the framework of the present analysis, it is not known
how the pumped energy is divided between the actual accumulation of
the splitting potential and absorption into a change of the
unperturbed 2-soliton's energy -- see Eq. (\ref{E0}) -- due to a
possible small
variation of $A$. While an exact prediction of $|p|$ as a function of $%
\epsilon $ seems to be too difficult for the fully analytical treatment, the
fact that, according to Fig. \ref{PowerLaw}, $|p|$ is essentially larger
than $1$ at $-\epsilon >10^{-3}$ suggests that, in this range of values of $%
\epsilon $, nearly all the pumped energy is absorbed into the increase of
the 2-soliton's energy. Combining Eqs. (\ref{E0}) and (\ref{E(t)}), one can
conclude that, as long as the resulting deviation of amplitude $A$ from its
initial value, $A_{0}$, remains small, the amplitude varies in time as $%
A(t)\approx A_{0}-\allowbreak 0.8A_{0}^{3}b\cdot t$. This variation leads to
a detuning of the resonant driving, according to Eq. (\ref%
{Eq_OscillationFrequency}), but, on the other hand, Fig. \ref{OffResonance}
demonstrates that the detuning does not produce an essential effect for $%
-\epsilon >10^{-3}$.

\section{Conclusions}

In this work we have considered the influence of the weak quintic
nonlinearity on the stability and splitting of second-order solitons (alias
2-solitons) in the NLSE-based model, under the action of the weak resonant
NLM (nonlinearity management), i.e., periodic time modulation of the
coefficient in front of the cubic term, with the frequency equal or close to
the frequency of the free shape oscillations of the 2-soliton. The model
applies to the propagation of light in CQ optical media, taking into regard
the periodic action of the linear loss and compensating gain. The same model
finds a natural application to BEC, where the self-focusing quintic term
accounts for the effect of the residual three-dimensionality in the
effective one-dimensional approximation, while the NLM may be induced by the
Feshbach resonance controlled by an ac magnetic field. By means of direct
simulations and an approximate analytical considerations, we have
demonstrated that the additional weak self-focusing quintic nonlinearity
stabilizes the 2-soliton, while the self-defocusing nonlinearity of the same
type makes it fragile and accelerates its splitting. We have confirmed the
resonant character of the splitting of the 2-soliton under the action of the
NLM, and proposed an explanation to this effect, based on the consideration
of the rate at which the energy is pumped into the bound state by the
resonantly tuned ac drive. We have also studied the resonant NLM-induced
splitting of the 2-soliton in the presence of the weak quintic nonlinearity.
Depending on its sign, the self-defocusing/focusing higher-order
nonlinearity gives rise to conspicuous broadening/sharpening of this
resonant response. The results of the numerical and analytical
considerations reported in this paper for 2-solitons can be readily extended
to higher-order solitons.


\begin{thebibliography}{99}
\bibitem{management} B. A. Malomed, \textit{Soliton Management in Periodic
Systems} (Springer: New York, 2006).

\bibitem{TheorySolitons} S. Novikov, S. V. Manakov, L. P. Pitaevskii, and V.
E. Zakharov, \textit{Theory of Solitons} (Consults Bureau, New York, 1984).

\bibitem{SolitonsInOC} G. P. Agrawal, \textit{Nonlinear Fiber Optics}
(Academic Press: San Diego, 2001); A. Hasegawa and Y. Kodama, \textit{%
Solitons in Optical Communication} (Oxford University Press: New York, 2004).

\bibitem{Malomed93} B. A. Malomed, D. F. Parker, and N. F. Smyth, Phys. Rev.
E \textbf{48}, 1418 (1993).

\bibitem{Bauer95} R. G. Bauer, L. A. Melnikov, Opt. Commun. \textbf{115},
190 (1995).

\bibitem{Sysoliatin08} A. A. Sysoliatin, A. K. Senatorov, A. I. Konyukhov,
L. A. Melnikov, and V. A. Stasyuk, Opt. Express \textbf{15}, 16302 (2008).

\bibitem{Hasegawa91} A. Hasegawa and Y. Kodama, Phys. Rev. Lett. \textbf{66}%
, 161 (1991).

\bibitem{Pare} C. Par\'{e}, A. Villeneuve, P.-A. Belang\'{e}, and N. J.
Doran, Opt. Lett. 21, \textbf{459} (1996); C. Par\'{e}, A. Villeneuve, and
S. LaRochelle, Opt. Commun. \textbf{160}, 130 (1999).

\bibitem{Radik} R. Driben, B. A. Malomed, M. Gutin, and U. Mahlab, Opt.
Commun. \textbf{218}, 93 (2003).

\bibitem{Radik2} R. Driben, B. A. Malomed, and U. Mahlab, Opt. Commun.
\textbf{232}, 129 (2004).

\bibitem{GPE} L. P. Pitaevskii and S. Stringari, \textit{Bose--Einstein
Condensation} (Clarendon Press: Oxford, 2003; ISBN 0-198-50719-4).

\bibitem{Greece} P. G. Kevrekidis, G. Theocharis, D. J. Frantzeskakis and B.
A. Malomed, Phys. Rev. Lett. \textbf{90}, 230401 (2003).

\bibitem{Scripta} R. Grimshaw, J. He, and B. A. Malomed, Phys. Scripta
\textbf{53}, 385 (1996).

\bibitem{Sakaguchi04} H. Sakaguchi and B. A. Malomed, Phys. Rev. E \textbf{70%
}, 066613 (2004).

\bibitem{18} G. S. Agarwal and S. Dutta Gupta, Phys. Rev. A \textbf{38},
5678 (1988).

\bibitem{19} E. L. Falc\~{a}o-Filho, C. B. de Ara\'{u}jo, and J. J.
Rodrigues, Jr., J. Opt. Soc. Am. B \textbf{24}, 2948 (2007).

\bibitem{20} R. A. Ganeev, M. Baba, M. Morita, A. I. Ryasnyansky, M. Suzuki,
M. Turu, and H. Kuroda, J. Opt. A: Pure Appl. Opt. \textbf{6}, 282 (2004).

\bibitem{ferro} B. Gu, Y. Wang, W. Ji, and J. Wang, Appl. Phys. Lett.
\textbf{95, }041114 (2009).

\bibitem{21} K. Dolgaleva, R. W. Boyd, J. E. Sipe, Phys. Rev. A \textbf{76},
063806 (2007).

\bibitem{22} F. Smektala, C. Quemard, V. Couderc, A. Barth\'{e}\'{l}\'{e}my,
J. Non-Cryst. Solids \textbf{274}, 232 (2000); K. Ogusu, J. Yamasaki, S.
Maeda, M. Kitao, and M. Minakata, Opt. Lett. \textbf{29}, 265 (2004); C.
Zhan, D. Zhang, D. Zhu, D. Wang, Y. Li, D. Li, Z. Lu, L. Zhao, and Y. Nie,
J. Opt. Soc. Am. B \textbf{19}, 369 (2002); G. Boudebs, S. Cherukulappurath,
H. Leblond, J. Troles, F. Smektala, and F. Sanchez, Opt. Commun. \textbf{219}%
, 427 (2003).

\bibitem{Shlyap02} A. E. Muryshev, G. V. Shlyapnikov, W. Ertmer, K.
Sengstock, and M. Lewenstein, Phys. Rev. Lett. \textbf{89}, 110401 (2002);
S. Sinha, A. Y. Cherny, D. Kovrizhin, and J. Brand, \textit{ibid}. \textbf{96%
}, 030406 (2006).

\bibitem{Luca} L. Salasnich, Laser Phys. \textbf{12}, 198 (2002); L.
Salasnich, A. Parola, and L. Reatto, Phys. Rev. A \textbf{65}, 043614 (2002).

\bibitem{Khaykovich06} L. Khaykovich and B. A. Malomed, Phys. Rev. A \textbf{%
74}, 023607 (2006).

\bibitem{Fatkhulla} Yu. Kagan, A. E. Muryshev, and G. V. Shlyapnikov, Phys.
Rev. Lett. \textbf{81}, 933 (1998); F. Kh. Abdullaev, A. Gammal, L. Tomio, \
and T. Frederico, Phys. Rev. A \textbf{63}, 043604 (2001).

\bibitem{Satsuma74} J. Satsuma and N. Yajima, Progr. Theor. Phys. Suppl.
\textbf{55}, 284 (1974).

\bibitem{Pushkarov79} Kh. I. Pushkarov, D. I. Pushkarov, and I. V. Tomov,
Opt. Quant. Electr. \textbf{11}, 471 (1979); S. Cowan, R. H. Enns, S. S.
Rangnekar, and S. S. Sanghera Can. J. Phys. \textbf{64}, 311 (1986).
\end{thebibliography}
\end{document}